\documentstyle{article} 
\newcommand{\be}{\begin{equation}}
\newcommand{\ee}{\end{equation}} 
\newcommand{\eei}{\end{equation}\indent\indent}
\newcommand{\bc}{\begin{center}}
\newcommand{\ec}{\end{center}}
\newcommand{\ber}{\begin{eqnarray}}
\newcommand{\ear}{\end{eqnarray}}
\newcommand{\ba}{\begin{array}}
\newcommand{\ea}{\end{array}}
\newcommand{\n}{{}^{(3)}\nabla} 

\newcommand{\hs}{\,-\,}

\def\case#1/#2{\textstyle\frac{#1}{#2} }
\begin{document}
\title{Covariant Analysis of Gravitational Waves\\
in a Cosmological Context}
\author{Peter K. S. Dunsby$^{1,2}$, Bruce A. C. C. Bassett$^3$
and George F. R. Ellis$^1$\\
$1.$ Department of Mathematics and Applied Mathematics,\\
University of Cape Town, South Africa.\\
$2.$ Department of Mathematics, Statistics and Computing Science,\\
Dalhousie University, Halifax, Nova Scotia, Canada.\\
$3.$ Scuola Internazionale Superiore de Studi Avanzati,\\
Strada Costiera 11, Miramare, 34014, Italy.}
\date{\today} 
\maketitle
\begin{abstract}
The propagation of gravitational waves or tensor perturbations in a perturbed 
Friedmann\hs Robertson\hs Walker universe filled with a perfect fluid is 
re\hs examined. It is shown that while the shear and magnetic part of the 
Weyl tensor satisfy linear, homogeneous {\it second order} wave equations, 
for perfect fluids with a $\gamma$\hs law equation of state satisfying 
$\case{2}/{3}<\gamma<2$, the electric part of the Weyl tensor satisfies a 
linear homogeneous {\it third order} equation. Solutions to these equations are 
obtained for a flat Friedmann\hs Robertson\hs Walker background and we discuss 
implications of this result.
\end{abstract}
\section{Introduction} \label{sec:standard}
Both the covariant definition and behaviour of gravitational waves and the 
dual question of the physical interpretation of the electric ($E_{ab}$) and 
magnetic ($H_{ab}$) parts of the Weyl tensor are subjects of 
debate at present. In particular the question arises whether it is possible 
to neglect the magnetic part of the Weyl tensor in the Newtonian limit, and 
whether this is equivalent to neglecting the gravitational wave contribution 
to cosmological perturbations. It is the aim in this paper to extend this 
debate by comparing the analysis of gravitational waves, set in a 
cosmological context, using the covariant approach on the one hand and 
the more standard methods based on metric perturbations on the other.

In the {\it standard approach} a second order propagation equation for the 
amplitude of the tensor perturbations is derived either via the variation 
of the Einstein\hs Hilbert action after expansion to second order, or by 
directly linearizing the Einstein field equations. For example in the 
Bardeen formalism \cite{bi:bardeen} the following equation of motion
is obtained for the first order gauge\hs invariant amplitude of the tensor 
perturbation $H_T^{(2)}$, in the absence of anisotropic stress perturbations:
\footnote{The equation for tensor metric perturbations was first derived by 
Lifshitz in 1946 \cite{bi:lifshitz}.}
\be
\ddot{H}_T^{(2)}+\case{2\dot{\ell}}/{\ell}\dot{H}_T^{(2)}+(k^2+2K)
H^{(2)}_T=0\;,
\label{eq:bard}
\ee
where $\ell$ is the cosmological scale factor, $K$ is
the spatial curvature of the background Friedmann\hs Robertson\hs Walker (FRW)
model, and $k$ is the physical wavenumber when $K=0$ \cite{bi:harrison}. 
The full metric tensor perturbation is $H_T^{(2)}Q_{\mu\nu}^{(2)}$, where 
$Q^{(2)}{}_{\mu\nu}$ are eigenfunctions (polarization tensors) of the tensor 
Helmholtz equation on the background spatial sections. These eigenfunctions 
have only two degrees of freedom after imposing the transverse 
($Q^{(2)\mu\nu}{}_{;\nu}= 0$) and traceless ($Q^{(2)\mu}{}_{\mu}=0$) 
conditions.

By contrast, the {\it covariant approach} focuses on either the electric 
or magnetic parts of the Weyl tensor \cite{bi:hawking,bi:ellis1}, 
and their propagation equations when linearized about a FRW model 
(which include waves propagating in a flat\hs space background as a limiting 
case). In this paper we will present a rather unusual result: while in a 
vacuum both $E_{ab}$ and $H_{ab}$ satisfy second order wave equations, when 
the spacetime is non\hs empty, and the matter is described by a barotropic 
perfect fluid with the equation of state $p=(\gamma-1)\mu$ satisfying 
$\case{2}/{3}<\gamma<2$, we obtain a third order equation for $E_{ab}$ but a 
second order equations for both $H_{ab}$ and the shear tensor $\sigma_{ab}$. 
This is somewhat like considering Maxwell's equations with sources 
\cite{bi:EH}, except in the gravitational case, the fundamental 
difference is that the source can never be turned off if there is matter 
present.

This paper is organized as follows: in section 2 we outline the covariant
characterization of gravitational waves. In section 3 we present differential
equations in closed form which describe the propagation of gravitational waves
on a FRW background with a perfect fluid source. Solutions to these
equations for a flat background are given in section 4 and the paper ends
with a discussion of the cosmological context of our results. 
\subsection{Notation and Conventions}
Conventions are taken to be the same as in \cite{bi:ellis1} and we
take $c=1$. It is assumed that Einstein's equations take the form 
$G_{ab}=\kappa T_{ab}$ where $G_{ab}$ is the Einstein tensor, $\kappa=8\pi G$ 
is the gravitational constant and $T_{ab}$ is the energy\hs momentum tensor 
of the matter, which is taken to be a perfect fluid. We use the standard
kinematical decomposition of the first covariant derivative of $u^a$ 
\cite{bi:ehlers,bi:ellis1}, a ``dot'' denotes the covariant derivative 
along the fluid flow lines and $\n_a$ corresponds to the 3\hs dimensional
covariant derivative defined by totally projecting the covariant derivative
orthogonal to $u^a$, so for example
\be
\dot{T}_{ab}=T_{ab;c}u^c\;,~~~\n_aT_{bc}=h^s{}_ah^t{}_bh^u{}_cT_{tu;s}\;,
\ee
where
\be
h_{ab}=g_{ab}+u_au_b
\ee
is the usual 3\hs dimensional projection tensor orthogonal to $u^a$.

A Robertson-Walker background geometry is characterized by the shear,
vorticity, and acceleration, together with their covariant derivatives, being 
at most first order, while the energy density and expansion are zero order, 
and the pressure can be so also \cite{bi:EB}. It follows that the electric 
and magnetic parts of the Weyl tensor:
\be
E_{ab}=C_{acbd}u^cu^d\;~~~~H_{ab}=\case{1}/{2}C_{acst}\eta^{st}{}_{bd}
u^cu^d
\ee 
are also at most first order. We will 
assume that in a cosmological context, $(\mu + p) \Theta > 0$. 

\section{The covariant approach to gravitational waves}
In the covariant approach, the study of tensor perturbations was first 
considered by Hawking \cite{bi:hawking}. In his paper, the electric part
of the Weyl tensor was used as the variable to characterize them. Later it 
was suggested that the Magnetic part of the Weyl tensor is a better choice 
because it has no analogue in Newtonian theory where gravity is propagated
instantaneously. Hence the Magnetic part obviously plays an important role
in describing gravitational waves, but given the correspondence with
electromagnetism \cite{bi:EH}, where neither the electric nor magnetic
fields provide a complete description of EM waves, we suggest that both
electric and magnetic parts of the Weyl tensor are required for a full
understanding of tensor perturbations. Indeed it is their {\it curls} that 
characterize gravitational waves, as we will see below. 

The fully {\it nonlinear} evolution equations for $E_{ab}$ and $H_{ab}$, with 
the matter source described by a perfect fluid, are given by \cite{bi:ellis1}:
\ber
\lefteqn{h^{m}{}_a h^{t}{}_c\dot{E}^{ac}+h^{(m}{}_a\eta^{r)tsd}u_r 
H^{a}{}_{s;d}-2H{}^{(t}{}_q\eta^{m)bpq}u_b\dot{u}_p+\Theta E^{mt}}\nonumber\\
&&+h^{mt}(\sigma^{ab}E_{ab})-3E^{(m}{}_s\sigma^{t)s}-E^{(m}{}_s\omega^{t)s}
=-\frac{1}{2}(\mu+p)\sigma^{tm}
\label{eq:nle}
\ear
and
\ber
\lefteqn{h^{ma}h^{tc}\dot{H}_{ac}-h_{a}{}^{(m}\eta^{r)tsd}u_rE^{a}{}_{s;d}+ 
2E^{(t}_q\eta^{m)bpq}u_b\dot{u}_p}\nonumber\\
&&+h^{mt}(\sigma^{ab}H_{ab})+\Theta H^{mt}-3 H^{(m}{}_s\sigma^{t)s}
-H^{(m}{}_s\omega^{t)s}=0\;.
\label{eq:nlh}
\ear
Notice that just as in the electromagnetic case, the only difference between 
these equations is in the sign of the second and third terms, and the source 
term (here, the shear) coupled to the energy density and pressure in the 
$\dot{E}_{ab}$ equation. Once linearized about a FRW background, these 
equations become: 
\be
\dot{E}_{ab}+\Theta E_{ab}+\n^eH^d{}_{(a}\eta_{b)cde}u^c
+\case{1}/{2}\kappa(\mu+p)\sigma_{ab}=0 \label{eq:dotE}
\ee
and 
\be
\dot{H}_{ab}+\Theta H_{ab}-\n^eE^d{}_{(a}\eta_{b)cde}u^c=0\;.
\label{eq:dotH}
\ee
On taking the time-derivative of the first and substituting from
the second, it is the {\it curl} $E$ and {\it curl} $H$ terms that give 
rise to traveling gravitational waves, in analogy with the propagation of 
electromagnetic waves.

It is immediately clear from the above that (given the FRW background 
evolution, which determines the zero-order coefficients in the equations), 
these equations by themselves do not close up in general, because of
the last term in (\ref{eq:dotE}). One has to add 
to them the shear evolution equation, which in linearized form is
\begin{equation}
\dot{\sigma}_{ab}=\n_{(a}\dot{u}_{b)}-\case{1}/{3}h_{ab} 
\dot{u}^{c}{}_{;c}-\case{2}/{3}\Theta\sigma_{ab}-E_{ab}\;.
\label{eq:shear}
\end{equation}
This is crucial in leading to the third\hs order equation for $E_{ab}$, on
taking the time derivative of (\ref{eq:dotE}).

One must also be aware of the constraint equations that have to be satisfied. 
In particular, after linearization we find the `div E' and `div H' equations
\ber
E^{ab}{}_{;b}&=&\case{1}/{3}\kappa X^a 
\label{eq:divE}\\
H^{ab}{}_{;b}&=&(\mu+p)\omega^a\;, 
\label{eq:divH}
\ear
where $X^a\equiv h^{ab}\mu_{,b}$ is the spatial gradient of the energy density 
\cite{bi:EB} and $\omega^a$ is the vorticity vector, 
which are the remaining `Maxwell-like' equations for $E_{ab}$ and
$H_{ab}$. 
\section{Closed evolution equations for linear gravitational waves}
In linear perturbations of FRW models, {\it pure tensor} perturbations 
are characterized by requiring that the following two quantities vanish to
{\it first order} \cite{bi:hawking,bi:DE}:
\ber
X^a = 0 & \Rightarrow & E^{ab}{}_{;b} = 0\,,\label{eq:divE1} \\ 
\omega^a = 0 & \Rightarrow & H^{ab}{}_{;b} = 0\;,
\label{eq:divH1}
\ear
the first one excluding scalar (density) 
perturbations and the second one, vector (rotational) perturbations. 
The conditions that the terms on the right hand side vanish, following from
(\ref{eq:divE},\ref{eq:divH}), are analogous to the transverse condition on tensor 
perturbations in the metric approach, 
In addition, we notice that since the Weyl tensor is the trace\hs free 
part of the Riemann tensor, both $E_{ab}$ and $H_{ab}$ are trace\hs free, 
again like the tensor perturbations of the Bardeen approach. 

Given the assumed equation of state, these conditions also imply that the 
spatial gradients of the pressure and expansion and the 
acceleration vanish:
\be
Y_a\equiv h^b{}_ap_{,b}=0\;,~~~Z_a\equiv h^b{}_a\Theta_{,b}=0\;,
~~~\dot{u}_a=0\;,
\ee
the first following from the equation of state, the second from the
evolution equations for $X^a$ (see \cite{bi:EB}), and the third from the
momentum conservation equations. Together with (\ref{eq:divE},\ref{eq:divH}) 
these characterize tensor perturbations. 

Once conditions
(\ref{eq:divE1}) and (\ref{eq:divH1}) have been imposed, one might suspect 
that both $E_{ab}$ and $H_{ab}$ would show the usual symmetry, i.e. that 
their propagation equations are the same under the substitution $E_{ab} 
\Leftrightarrow H_{ab}$. However as we will see below, this is not the case 
in general. In fact, we will demonstrate that it is only true when the 
equation of state of the background spacetime satisfies very special 
conditions.
In general, as mentioned in the introduction, the equations for $E_{ab}$
and  $H_{ab}$ are not even of the same {\em order}, the former satisfies a
third order equation while the latter satisfies a second order one. In order 
to see this, let us begin by considering the linearized second\hs order 
equation for tensor perturbations, obtained by taking the time derivative 
of (\ref{eq:dotE}) and substituting from (\ref{eq:dotH}):
\be
\Delta E_{ab}+\case{7}/{3}\Theta\dot{E}_{ab}+\left(\case{2}/{3}\Theta^2
-2p\right) E_{ab}-\case{1}/{6}\mu\gamma\left[\left(3\gamma-2\right)\Theta
-\case{3\dot{\gamma}}/{\gamma}\right]\sigma_{ab} = 0\;,
\label{eq:edd}
\ee
where $\Delta E_{ab}=\ddot{E}_{ab}-\n^2E_{ab}$. By differentiating this 
equation, using the linearised shear evolution equation 
(\ref{eq:shear}) specialized to tensor perturbations:
\be
\dot{\sigma}_{ab}=-\case{2}/{3}\Theta\sigma_{ab}-E_{ab}\;, 
\label{eq:shear2}
\ee
to eliminate $\dot{\sigma}_{ab}$, and using (\ref{eq:edd}) again to eliminate 
the term in $\sigma_{ab}$, 
we can eliminate the remaining shear dependence, however this still doesn't 
give an equation that is closed. This situation can be rectified if we 
perform a harmonic decomposition, expanding $E_{ab}$ in terms of 
eigenfunctions 
of the Laplace\hs Beltrami operator of the background FRW model: $E_{ab}=\sum 
E_{(k)} Q^{(k)}{}_{ab}$. In this way we can replace the Laplacian term 
$^{(3)}\nabla^2$ in (\ref{eq:edd}) by $k^2/\ell^2$ where 
$k$ is the physical wave number if the background is flat \cite{bi:harrison}. 
This gives us the closed, third order evolution equation for the harmonic 
components of $E_{ab}$:
\ber
\stackrel{\ldots}{E}_{(k)}&+&\left[3\Theta-\case{\dot{B}}/{B}\right]
\ddot{E}_{(k)} 
+\left[\case{7}/{9}\Theta^2-\case{7}/{6}\left(\mu+3p\right)+A-\case{7}/{3}
\case{\Theta\dot{B}}/{B}\right]\dot{E}_{(k)}\nonumber\\
&+&\left[\dot{A}-\case{A\dot{B}}/{B}+\case{2}/{3}\Theta A-B\right]E_{(k)}=0\;,
\label{eq:third}
\ear
where         
\be
A=\case{2}/{3}\Theta^2-2p+\case{k^2}/{\ell^2}
\ee
and
\be
B=-\case{1}/{6}\mu\gamma\left[\left(3\gamma-2\right)\Theta
-\case{3\dot{\gamma}}/{\gamma}\right]\;.
\ee
In contrast, the propagation equation for $H_{ab}$ is relatively simple: 
\be
\Delta H_{ab}+\case{7}/{3}\Theta\dot{H}_{ab}+\left(\case{2}/{3}\Theta^2
-2p\right)H_{ab}=0\;. 
\label{eq:hdd}
\ee
The reason why the equation for $H_{ab}$ is second order is due to 
the constraint equation \cite{bi:ellis1} (with vanishing vorticity):
\be
H_{ab}=-\n^e\sigma^d{}_{(a}\eta_{b)cde}u^c\;.
\label{eq:const}
\ee
This allows the shear term - arising on taking the time derivative of 
(\ref{eq:dotH}) and substituting from (\ref{eq:dotE}) - to be replaced 
in terms of $H_{ab}$. In the case of 
the equation for $E_{ab}$ (\ref{eq:edd}), the shear term cannot be removed  
without differentiating again, leading to the higher order equation. 

The shear also satisfies a second order equation. This follows by 
differentiating equation (\ref{eq:shear2}) and using (\ref{eq:dotE}) and 
(\ref{eq:const}) to substitute for $E_{ab}$ and $H_{ab}$:
\be
\Delta\sigma_{ab}+\case{5}/{3}\Theta\dot{\sigma}_{ab}+\left[\case{1}/{9}
\Theta^2+\case{1}/{6}\mu\left(9\gamma-1\right)\right]\sigma_{ab}=0\;.
\ee

By comparing the equations (\ref{eq:edd}) for $E_{ab}$ and (\ref{eq:hdd})
for $H_{ab}$, it is clear that they will be identical in form iff the 
equation of state (assumed to satisfy $\gamma \mu \neq 0$) satisfies
\be
\dot{\gamma}-\case{1}/{3}\gamma\left(3\gamma-2\right)=0\;,
\label{eq:cond}
\ee
for then the last term in (\ref{eq:edd}) will vanish.  
In the case of a barotropic perfect fluid where $\gamma$ is a constant, two
possibilities exist: either $\gamma=0\Leftrightarrow \mu+p=0$ corresponding 
to a false vacuum, already rejected, or $\gamma=\case{2}/{3}$ which represents 
the coasting solution and is also non-physical. If $\gamma$ is allowed to 
vary, then (\ref{eq:cond}) will also be satisfied provided
\be
\gamma=\frac{2}{\beta-\exp{\case{2}/{3}t}}\;, \label{eq:vary}
\ee
where $\beta$ is an arbitrary constant. However this will usually have to 
be rejected because it will then, for early or late times, exceed the standard 
bounds $0 \leq dp/d\mu \leq 1$ demanded for physical reasons \cite{bi:ellis1} 
(note that when $\gamma$ varies, it is no longer true that $dp/d\mu = \gamma$).
For normal matter satisfying $\case{2}/{3}<\gamma<2$, the equations for 
$E_{ab}$ and $H_{ab}$ will be very different.

The appearance of a third order equation for the electric part of the Weyl 
tensor $E_{ab}$ is very surprising. First of all the standard theory 
discussed in section \ref{sec:standard} gives a second order evolution 
equation (\ref{eq:bard}), and secondly, force laws are expected to be 
formulated as second order evolution equations. We consider this again in
section 6. Hawking's covariant analysis 
\cite{bi:hawking} gave a second order equation, because he assumed a vacuum
condition, instead of consistently sticking to a cosmological context (as
assumed here). We consider this again in section 5.1.

\section{Solutions for a flat FRW background}
At first glance there seems to be an inconsistency with waves mediated by 
$E_{ab}$ and $H_{ab}$, where one is governed by a second order equation
and the other by a third order equation. However consistency of the equations 
has been carefully checked \cite{bi:DE,bi:HE}. We need then 
to explore the nature of the solutions, and also check whether there 
is some hidden symmetry in the equations that makes them compatible. 

For simplicity we will solve the above equations only in the case of a 
flat FRW background ($K=0$). Using the conformal time variable $\eta$: 
$\case{d\eta}/{dt}=\case{1}/{\ell}$, the equations for the harmonic 
components of the shear and magnetic part of the Weyl tensor become
\be
 \sigma''_{(k)}+\case{4\ell'}/{3\ell}\sigma'_{(k)}-\left[\case{1}/{2}
\mu\ell^2\left(3\mu-4\right)-k^2\right]\sigma_{(k)}=0\;,
\label{eq:main1}
\ee
and
\be
H''_{(k)}+\case{6\ell'}/{\ell}H'_{(k)}-\left[2\mu\ell^2\left(\gamma-2\right)
-k^2\right] H_{(k)}=0\;.
\label{eq:main2}
\ee
for a flat background the scale factor $\ell$, energy density $\mu$ and 
expansion $\Theta$ are given by:
\be
\ell\propto\eta^{\case{2}/{3\gamma-2}}\;,~~~
\mu\propto\eta^{-\case{6\gamma}/{3\gamma-2}}\;,~~~
\Theta\propto\eta^{-\case{3\gamma}/{3\gamma-2}}\;.
\ee
Substituting the background parameters into (\ref{eq:main1}) 
and (\ref{eq:main2}), and integrating, gives the following general solutions
\be
\sigma_{(k)}=\eta^{\case{3\gamma-10}/{2(3\gamma-2)}}\left[J_{\nu}(k\eta)
C^{(1)}_{(k)}+N_{\nu}(k\eta)C^{(2)}_{(k)}\right]\;,
\ee
and
\be
H_{(k)}=\eta^{\case{3\gamma-14}/{2(3\gamma-2)}}\left[[J_{\nu}(k\eta)
D^{(1)}_{(k)}+N_{\nu}(k\eta)D^{(2)}_{(k)}\right]\;,
\ee
where $C$ and $D$ are arbitrary constants and $J_{\nu}$, $N_{\nu}$ denote
the Bessel functions of the first and second kind. 

For large scales ($k\rightarrow 0$), we obtain power\hs law solutions:
\be
\sigma_{(k)}=\eta^{\case{3\gamma-4}/{3\gamma-2}}C^{(1)}_{(k)}
+\eta^{-\case{6}/{3\gamma-2}}C^{(2)}_{(k)}\;.
\ee
and 
\be
H_{(k)}=D^{(1)}_{(k)}\eta^{\case{3(\gamma-2)}/{3\gamma-2}}
+D^{(2)}_{(k)}\eta^{-\case{8}/{3\gamma-2}}\;.
\ee
In the case of a flat background, the third order equation for $E_{(k)}$ 
(\ref{eq:third}) becomes: 
\ber
E'''_{(k)}&+&\left(4+3\gamma\right)\case{\ell'}/{2\ell}E''_{(k)}
+\left[(4\gamma+6)\mu\ell^2+k^2\right]E'_{(k)}\nonumber\\
&+&\left\{\case{1}/{6}\left[16-\gamma\left(3\gamma-2\right)\right]\mu\ell^2
+\case{3}/{2}\gamma k^2\right\}\ell\Theta E_{(k)}=0\;.
\ear
This time, a general solution could not be found,  
however for large scale modes we obtain:
\be
E_{(k)}=\eta^{-\case{3(\gamma+2)}/{3\gamma-2}}E^{(1)}_{(k)}
+\eta^{-\case{4}/{3\gamma-2}}E^{(2)}_{(k)}
+\eta^{\case{3\gamma-8}/{3\gamma-2}}E^{(3)}_{(k)}\;.
\ee
The first two of these modes were previously known \cite{bi:goode1,bi:goode2}. 
The third one, due to the third order nature of the equation for 
$E_{(k)}$, has to be unphysical as it is not a solution of the original 
system of first order equations. The point is an essential one: 
one cannot satisfy one or other of these equations in isolation, one has to 
solve the set as a whole. In this case, we can solve first for $\sigma_{ab}$ 
(a second order equation), as above, and then determine $E_{ab}$ from 
(\ref{eq:shear2}) by simply differentiating this solution. This will 
determine solutions to the third order equation for $E_{ab}$; but there 
will be two modes, not three:
\be
E_{(k)}=\eta^{-\case{3(\gamma+2)}/{3\gamma-2}}E^{(1)}_{(k)}
+\eta^{-\case{4}/{3\gamma-2}}E^{(2)}_{(k)}\;.
\ee
This shows that in fact the effective equation governing the solution for
$E_{ab}$, considered as a solution to the set of equations as a whole, is 
second order, corresponding to the previously known solutions, even though 
one apparently cannot write down a separate second order equation for 
$E_{ab}$ alone.
\section{The cosmological context}
We have emphasized here that we are considering the problem of gravitational
waves in a cosmological setting.
\subsection{Averaging}
The key point in the analysis has been a non-vacuum assumption. Is this 
realistic?

As discussed above, the evolution equations for $E_{ab}$ and $H_{ab}$ are 
the same when $\mu+p=0$. Now for astrophysical sources of gravitational 
waves, such as pulsars or merging binary systems in our own 
galaxy, the vacuum assumption is likely to be good for much of the region 
the wave travels in, since the intergalactic and interstellar media are
relatively tenuous. Waves with wavelength less than 1 Mpc will in fact 
travel through a vacuum for much of the time; and those of wavelengths less 
than 1 pc will be in an effective vacuum for almost all the time. 
Thus in this case the Hawking analysis will be adequate; both $E_{ab}$
and $H_{ab}$ will satisfy the same second-order wave equation.

However cosmological scale gravitational waves, and certainly those with 
wavelengths of the order of the horizon size, will experience a non\hs 
vacuum spacetime background all the time. Thus our problem is one of 
averaging: on what scale does the geometry of the universe start to 
approximate that of a Robertson\hs Walker universe? \cite{bi:ES} On 
smaller scales, the vacuum approximation will be acceptable for much but 
not all the time; on larger scales it is not, and the third order equation
discussed above for $E_{ab}$ applies.

In particular, we can ask the following question: What happens to $E_{ab}$ 
for small-scale waves if, after traveling in vacuum, it comes across a 
non\hs zero matter concentration, for example in the case of a gravitational 
wave crossing a galaxy cluster? The wave equation for the waves will
change then from second to third order. However perhaps this is not too
drastic, as the effective order of the set of equations will remain second 
order, as discussed above. Nevertheless the junction conditions at such a 
change need careful consideration, because the standard assumptions allow 
a discontinuity of the curvature tensor at a boundary of this kind. We will
not pursue this further here, but note it as worthy of investigation. 

\subsection{Sharp phase transitions}\label{sec:phase}
The same kind of issue arises in the early universe, in a different context.

If the early universe went through an inflationary phase, then there 
must have been a transition at the end of inflation from the de\hs Sitter 
phase to a radiation dominated phase, often treated as an instantaneous 
3\hs surface, although in reality reheating is not instantaneous. The 
question we pose is as follows: can there be amplification of perturbations 
across this boundary? The standard answer is that density perturbations can be 
amplified 
, but that tensor perturbations cannot. This is 
simply a result of the Darmois junction conditions which require that the 
three\hs metric and the extrinsic curvature of the three\hs metric be 
continuous across the boundary. 

The key point for us is that for tensor perturbations, this theorem of no\hs 
amplification does not carry through in the covariant case since both the 
electric and magnetic parts of the Weyl tensor will depend on the second 
derivatives of the metric perturbations, which are unconstrained by the 
junction conditions. Thus we cannot agree with the conclusion that 
gravitational waves cannot be amplified across a sharp phase transition. 
We will not discuss this issue further here; it is raised because it
is similar to the situation just discussed.

\section{Conclusion}
It is interesting that the evolution equations for $E_{ab}$ and $H_{ab}$ 
are so different, because this implies a radical break with the analogy of 
source-free electromagnetism, which had been thought to carry through almost 
completely, at least qualitatively. This is assuming of 
course that the evolution equation for $E$ cannot be reduced to second order
(the `hidden symmetry' option). We have not found a way to do so; the third
order equation appears to us to be genuinely third order. 

Clearly it is interesting to locate the conditions when 
the evolution equations for $E_{ab}$ and $H_{ab}$ reduce to the same 
second order equation. We found that this happens (i) when $(\mu + p) = 0$, 
i.e. in vacuum and in de Sitter spacetime, (ii) for the
exceptional equation of state $p = - (1/3)\mu$, (iii) and the unphysical
variation of $\gamma$ given by (\ref{eq:vary}). We exclude all these cases
in a realistic cosmological context, for long-wavelength waves. 

Our analysis shows that for large scale gravitational waves in a universe 
model with realistic matter, $E_{ab}$ and $H_{ab}$ obey quite different 
equations, yet these are consistent with each other. 
This is like the situation for Maxwell's equations with a source, as is 
shown in the accompanying paper \cite{bi:EH}. Although the wave equation for
$E_{ab}$ is third order, it has the same characteristics as the second order
equation for $H_{ab}$, and both can be solved from a single `potential',
namely the shear (obeying a second order wave equation). This presumably 
is how the Bardeen analysis leads only to a second order equation: that 
equation is also for a potential, and the equation satisfied by the Weyl 
tensor itself is not written down. 


\end{document}